\documentclass[11pt]{article}
\usepackage{epsfig,sint,cite,amsfonts}

\usepackage[english]{babel}
\usepackage{simplewick}

\newcommand{\be}{\begin{equation}}
\newcommand{\ee}{\end{equation}}
\newcommand{\bea}{\begin{eqnarray}}
\newcommand{\eea}{\end{eqnarray}}

\newcommand{\Tr}{\,\hbox{\rm Tr}}

\newcommand{\ranglec}{{\rangle_c}}

\newcommand{\bp}{{\bf p}}
\newcommand{\Fgen}{K}
\newcommand{\bz}{{\bf z}}
\newcommand{\bx}{{\bf x}}

\newcommand{\bm}{{\bf m}}
\newcommand{\<}{\langle}
\renewcommand{\>}{\rangle}

\begin{document}

\begin{titlepage}

\begin{center}
\begin{flushright}
\end{flushright}
\vspace{2.5cm}

{\Large\bf Thermodynamic potentials from shifted boundary\\[0.15cm]
conditions: the scalar-field theory case\\[0.5ex]} 

\end{center}
\vskip 0.5 cm
\begin{center}
{\large  Leonardo Giusti$^{\scriptscriptstyle a}$ and Harvey B. Meyer$^{\scriptscriptstyle b}$}
\vskip 0.75cm
$^{\scriptstyle a}$ Dipartimento di Fisica, Universit\'a di Milano-Bicocca,\\ 
                    Piazza della Scienza 3, I-20126 Milano, Italy\\
\vskip 1.5ex
$^{\scriptstyle b}$ Institut f\"ur Kernphysik, University of Mainz,\\
                    Johann-Joachim-Becher Weg 45, D-55099 Mainz, Germany\\
\vskip 2.0cm
{\bf Abstract}
\vskip 0.35ex
\end{center}

\noindent
In a thermal field theory, the cumulants of the momentum distribution
can be extracted from the dependence of the Euclidean path integral on
a shift in the fields built into the temporal boundary condition. When 
combined with the Ward identities associated with the invariance of the theory 
under the Poincar\'e group, thermodynamic potentials such as the entropy or the 
pressure can be directly inferred from the response of the system to the shift. Crucially 
the argument holds, up to harmless finite-size and discretization effects, even if 
translational and rotational invariance are broken to a discrete subgroup of finite 
shifts and rotations such as in a lattice box. The formulas are thus applicable at finite 
lattice spacing and volume provided the derivatives are replaced by their discrete 
counterpart, and no additive or multiplicative ultraviolet-divergent renormalizations are 
needed to take the continuum limit. In this paper we present a complete derivation of the 
relevant formulas in the scalar field theory, where several technical complications are 
avoided with respect to gauge theories. As a by-product we obtain a recursion 
relation among the cumulants of the momentum distribution, and formul\ae~for 
finite-volume corrections to several well-known thermodynamic identities.
\vskip 4.5cm

\vfill

\eject

\end{titlepage}

\section{Introduction}
Thermal field theory is the theoretical tool for computing properties of 
matter at high temperatures and densities from first principles~\cite{Kapusta:2006pm,LeBellac}. 
It allows one, for instance, to determine the equation of state of 
Quantum Chromodynamics, which 
in turn is an essential ingredient to understand the properties of matter created in heavy ion 
collisions, and to model the behavior of hot matter in the early universe (for recent 
reviews see Ref.~\cite{Laine:2009ik,Philipsen:2010gj}). Obtaining first-principles 
predictions from a thermal field theory is often challenging since it describes an infinite
number of degrees of freedom subject to both quantum and thermal fluctuations.  Even though 
there are several established methods to compute the thermal properties of field 
theories~\cite{Engels:1981qx,Engels:1990vr,Endrodi:2007tq,Meyer:2009tq}, new theoretical 
concepts and more efficient computational techniques are still needed in many contexts
particularly when weak-coupling methods are inapplicable.

In a recent Letter we related the generating function of the cumulants 
of the total momentum distribution to a path integral with properly 
chosen shifted boundary conditions in the compact direction normalized 
to the ordinary thermal one~\cite{Giusti:2010bb}. By exploiting 
the Ward Identities (WIs) associated with the space-time invariances of 
the continuum theory, the cumulants can be related in a simple manner to 
thermodynamic potentials. In a relativistic theory at zero chemical potential, 
for instance, the variance of the momentum probability distribution 
measures the entropy of the system. Thanks to a recursion relation
among the cumulants, the kurtosis is related to the specific heat. 
These results suggested a new way to determine the equation of state
of a thermal field theory~\cite{Giusti:2010bb}.

The aim of this paper is to present a complete and self-contained 
derivation of the formulas introduced in Ref.~\cite{Giusti:2010bb} 
for the scalar field theory, where several technical complications 
in their derivation are avoided with respect to gauge 
theories~\cite{Giusti:2011}. A crucial r\^ole is played by the 
symmetry-constrained path 
integrals~\cite{DellaMorte:2008jd,DellaMorte:2010yp} and by 
the continuum WIs associated with the relativistic invariance of the theory. 
The latter allow us to derive the recursive relation among the cumulants, 
to generalize well-known thermodynamic relations to finite-volume systems, and to 
exclude additive or multiplicative ultraviolet-divergent renormalizations
of the cumulants. The formulas are applicable, up to harmless 
finite-size and discretization effects, at finite volume and lattice 
spacing, where ratios of path integrals can be determined by 
\emph{ab initio} Monte Carlo computations~\cite{Giusti:2010bb}. 
As a result the entropy density, the pressure and the specific heat of a 
thermal field theory can be obtained by studying the response of the system 
to the bare shift parameter. 

After a section on the basic properties of the scalar theory, the 
symmetry-constrained path integral and its relation with the cumulant 
generator is introduced in section \ref{sec:SyCPI}. The relevant WIs 
are derived in sections \ref{sec:wi1} to \ref{sec:wi3}, and the main results 
and conclusions of the paper are reported in section \ref{sec:mainres}. 
Several technical details are given in the four appendices.

\section{Preliminaries and basic notation\label{sec:prel}}
We are interested in the thermal scalar theory defined 
via the Euclidean path integral formalism in an infinite volume
as well as in a finite box of volume\footnote{Throughout the paper 
the linear dimension in the spatial direction $k$ will be indicated by $L_k$.} 
$V=L^3$ with the field $\phi$ satisfying periodic boundary conditions.
The time extent is set to $L_0=1/T$, where $T$ is the temperature of 
the system. Some basic definitions and WIs, associated with the 
invariance of the continuum theory under the Poincar\'e group, are 
reviewed in this section. Other properties of the theory, not 
directly relevant for the subject of this paper, can be found in various 
textbooks, see for instance Refs~\cite{ZinnJustin:1989mi,Kleinert:2001ax}.

The partition function 
of the theory is defined as usual as 
\be
Z = \int D \phi\; e^{-S}\; , 
\ee
where the action $S=\int d^4 x\, {\cal L}$ is defined by the Lagrangian density 
\be
{\cal L} = \frac{1}{2}\, (\partial_\mu\phi) (\partial_\mu\phi) + V(\phi)\; , \qquad
V(\phi) = \frac{1}{2}\, m^2 \phi^2 + \frac{\lambda}{4!}\, \phi^4\; .
\ee
The equations of motion are given by 
\be
\left\langle \left\{\frac{\partial V}{\partial\phi} - \Box\phi \right\}\!\!(x)\; 
O_1 \dots O_n\right\rangle = 
\sum_{i=1}^n\; \langle\, O_1 \dots O'_i  \dots O_n\, \rangle \; , 
\ee
where $O_i$ is a generic local field\footnote{Sometimes the argument 
$x^i$ is shown explicitly to clarify the meaning of some 
formulas. Summation over repeated indices is understood unless 
explicitly specified. No summation over $k$ is understood in sections 
\ref{sec:wi1} to \ref{sec:wi3}, and everywhere for the field $T_{kk}$.}
inserted at the point $x^i$, and $ O'_i$ is its variation 
with respect to the fundamental field and its derivative at the coordinate 
value $x$. 

\subsection{Translational invariance}
The theory is invariant under space-time translations, i.e. under 
the direct product of four continuum Abelian groups of global transformations 
\be
x_\mu' = x_\mu - \varepsilon_\mu\; , \qquad \phi'(x_\mu') = \phi(x_\mu)\; .  
\ee
The associated WIs can be derived in the usual way by promoting the symmetry 
to a local one, i.e. $\varepsilon_\mu \rightarrow \varepsilon_\mu(x)$, and 
studying the variation of the functional integral under the local 
transformations
\be\label{eq:trdelta}
\delta\phi(x) = \varepsilon_\mu(x)\, \partial_\mu \phi(x)\; , \qquad
\delta\{\partial_\nu \phi(x)\} = \partial_\nu\{\varepsilon_\mu(x) \partial_\mu \phi(x)\}\; .
\ee
By considering different functions $\varepsilon_\mu(z)$ and properly chosen fields 
$O_i$, many interesting non-trivial relations can be derived. For 
$\varepsilon_\nu(z) = \epsilon_\nu \delta^{(4)}(z-x)$ one obtains 
\be\label{eq:WIstd1}
\epsilon_\nu\, \langle \partial_\mu T_{\mu\nu}(x)\, O_1 \dots O_n \rangle = -  
\sum_{i=1}^{n}\, \left\langle O_1 \dots \delta^x_\epsilon O_i \dots O_n \right\rangle\; , 
\ee
where $\delta^x_\epsilon O_i$ is the variation of 
the field $O_i$ under the transformation (\ref{eq:trdelta}), and  
the field 
\be\label{eq:Tmunu}
T_{\mu\nu} = (\partial_\mu\phi) (\partial_\nu\phi) - 
\delta_{\mu\nu} {\cal L}
\ee
is the energy-momentum tensor of the theory symmetric under 
the exchange $\mu\leftrightarrow\nu$. The WIs~(\ref{eq:WIstd1}), and therefore 
the consequences discussed in the following, applies to connected correlation 
functions as well. When all operators $O_i$ are localized 
far away from $x$, the classical conservation identities  
\be\label{eq:WIstd2}
\langle \partial_\mu T_{\mu\nu}(x)\, O_1 \dots O_n \rangle = 0\;   
\ee
are recovered. If we integrate Eq.~(\ref{eq:WIstd1}) over a bounded 
region $R$ which contains in its interior the points $x^1,\dots,x^m$ while 
the fields $O_{m+1}\dots O_{n}$ are localized outside, the Gauss theorem 
leads to  
\be\label{eq:mixWI}
\hspace{-0.25cm}\epsilon_\nu \int_{\partial R} d \sigma_\mu(x)\, 
\langle T_{\mu\nu}(x) O_1 \dots O_n \rangle = -  
\sum_{i=1}^{m}\, \left\langle O_1 \dots \delta_\epsilon O_i \dots O_n \right\rangle\; , 
\ee 
where $\delta_\epsilon O_i$ is the variation of the field $O_i$ under 
the {\it global} transformation associated with $\epsilon_\nu$, and the integration 
measure $d \sigma_\mu(x)$ points along the outward 
normal to the surface $\partial R$. In this case, the very same WIs 
could have been obtained by considering directly the limit where 
$\varepsilon_\nu(x)$ goes to a constant in $R$, and it is 
null outside. 

The discussion above applies also to the theory defined in a finite box
with periodic boundary conditions, provided the $\delta$-function 
is replaced by its periodic generalization $\delta^{({\rm p})}$. In the following
we will use the same notation for both of them since the 
precise meaning will be clear from the context. 

\subsection{Time evolution of operators}
From the WIs in Eq.~(\ref{eq:mixWI}) we can derive the 
time-evolution of a generic field. If we choose 
$\epsilon_\nu = \delta_{\nu 0} \epsilon_0$, 
and we integrate
over a thick time-slice $R$ with the field $O_1$ being 
an arbitrary functional of $\phi$ and $\partial_\mu\phi$
inserted into it while the operators $O_2\dots O_n$ are localized 
outside, then 
\be
\delta O_1(x) = \epsilon_0\, \partial_0 O_1(x)
\ee
and therefore
\be
\partial_0 \, \langle  O_1(x^1)\, O_2\dots O_n\rangle
= - \int_{\partial R} d \sigma_0(x)\, \langle T_{00}(x)\, O_1(x^1)\, O_2\dots O_n\ \rangle\; .
\ee
This is the Euclidean version of the time-evolution of the generic field
$O_1(x^1)$. The field 
\be
{\overline T}_{00}(x_0) = \int d^3{\bf x}\, T_{00}(x)
\ee
is therefore (minus) the Hamiltonian of the system, and we can write 
\be
\langle {\overline T}_{00}\rangle = \frac{\partial}{\partial L_0} \ln{Z}\; ,
\ee
or more generally
\be\label{eq:devL0O}
\langle {\overline T}_{00}(L_0)\, O \rangle_c = 
\frac{\partial}{\partial L_0}\, \langle O \rangle\; , 
\ee
where $O$ is a generic operator which does not depend explicitly on the time 
coordinate and it is at a physical distance from the time-slice $L_0$.

\subsection{Translational invariance of correlators}
If all fields $O_1,\dots,O_n$ are inside the integration 
region $R$, then the transformation (\ref{eq:trdelta}) implies 
\be\label{eq:gtrans}
\delta O_i(x) = \epsilon_\mu\, \partial_\mu O_i(x)
\ee
for arbitrary functionals of $\phi$ and $\partial_\mu\phi$.
In the limit where the bounded region $R$ goes to infinity or, for instance,
in a finite volume with periodic boundary conditions, the l.h.s. of 
Eq.~(\ref{eq:mixWI}) can be neglected\footnote{In WIs associated
to space-time translational invariance no boundary terms of this type are expected 
to contribute at infinity in the quantum theory, at variance of what happen in those 
associated with spontaneously broken symmetries.}. Correlation functions enjoy translational 
invariance, i.e. they satisfy
\be
\sum_{i=1}^{n}\, \partial^{x^i}_\mu \;
\langle O_1(x^1) \dots  O_i(x^i) \dots O_n(x^n) \rangle =0 \; .
\ee
As an example of transformation of a composite field, it is interesting
to study the variation of the energy-momentum tensor under 
local transformations. If we  rewrite the tensor 
(\ref{eq:Tmunu}) as 
\be
T_{\mu\nu} = \frac{1}{2}\, R_{\mu\nu\alpha\beta}\, (\partial_\alpha\phi) (\partial_\beta\phi) 
- \delta_{\mu\nu} V(\phi)\; , \qquad R_{\mu\nu\alpha\beta} = 
\delta_{\mu\alpha}\delta_{\nu\beta} +
\delta_{\mu\beta}\delta_{\nu\alpha} -
\delta_{\mu\nu}\delta_{\alpha\beta}\; , 
\ee
its  variation under a local space-time translation (\ref{eq:trdelta}) can be written as  
\be
\delta T_{\mu\nu}(x) = \varepsilon_\rho(x)\, \partial_\rho T_{\mu\nu}(x) + 
                    R_{\mu\nu\alpha\beta} \left\{T_{\rho\beta}(x) + 
                    \delta_{\rho\beta} {\cal L}(x)\right\} 
                    \partial_\alpha\varepsilon_\rho(x)\; , 
\ee
and Eq.~(\ref{eq:gtrans}) is satisfied if $\varepsilon_\rho(x)$ is constant.

\subsection{Invariance under 4-dimensional rotations}
In the Euclidean the invariance under the homogeneous Lorentz 
group is replaced by the symmetry under $SO(4)$ rotations.
An infinitesimal transformation reads 
\be
x_\alpha' = x_\alpha + \omega_{\alpha\beta}\, x_\beta
\ee
with $\omega_{\alpha\beta} = - \omega_{\beta\alpha}$ and, as in the 
previous section, we can derive the associated WIs by studying 
the variation of the functional integral under an infinitesimal 
local transformation. The latter can be written as  
\be
x_\alpha' = x_\alpha - \varepsilon_\alpha(x)\; , \qquad
\varepsilon_\alpha(x) = - \omega_{\alpha\beta}(x)\, x_\beta\; , 
\ee
and the fields transform accordingly to Eq.~(\ref{eq:trdelta}).
If we choose $\omega_{\alpha\beta}(x) = w_{\alpha\beta} \delta^{(4)}(z-x)$,
we obtain
\be\label{eq:WIlrtz}
w_{\alpha\beta} \, \langle \partial_\mu K_{\mu;\alpha\beta}(x)\, 
O_1 \dots O_n \rangle = - 2
\sum_{i=1}^{n}\, \left\langle O_1 \dots \delta^x_w O_i \dots O_n  \right\rangle\; , 
\ee
where 
\be
K_{\mu;\alpha\beta} = x_\alpha T_{\mu\beta} - x_\beta T_{\mu\alpha} 
\ee
is an antisymmetric tensor for $\alpha\leftrightarrow\beta$. The 
WIs ~(\ref{eq:WIlrtz}) are just combinations 
of those in Eqs.~(\ref{eq:WIstd1}) since, thanks to the symmetry of 
$T_{\mu\nu}$, $\partial_\mu K_{\mu;\alpha\beta}$ is a linear combination of 
the components of $\partial_\mu T_{\mu\nu}$ with field-independent 
coefficients . All WIs associated with 4-dimensional rotations can thus 
be derived from those discussed in the previous section. It is, however, 
instructive to consider properly chosen integrated WIs. They can be 
obtained by integrating over a bounded region $R$ which contains in its interior 
the points $x^1,\dots,x^m$ while the fields $O_{m+1}\dots O_{n}$ are localized 
outside. This leads to  
\be\label{eq:WIlrtz2}
w_{\alpha\beta} \int_R d^4 x\, \langle \partial_\mu K_{\mu;\alpha\beta}(x)\, 
O_1 \dots O_n \rangle = - 2 \sum_{i=1}^{m}\, \left\langle O_1 \dots \delta_w O_i \dots O_n 
\right\rangle
\ee
where $\delta_w O_i$ is the variation of $O_i$ under the {\it global} transformations 
associated with $w_{\alpha\beta}$. In particular it is interesting to consider an 
infinitesimal boost, that in the Euclidean
is obtained by choosing $\omega_{0j}=v_j$ and all other components null, an   
integration region $R$ which is a thick-time slice between the two hyper-planes 
with $x_0=y_0 \pm t$, and a correlation function with 
a field ${\overline T}_{0k}(y_0)$ inserted inside $R$ while the 
operators $O_1\dots O_n$ are localized outside. The variation
of space-time components of the energy-momentum tensor 
under a global boost reads
\be
\delta T_{0k}(y) =   v_j \Big\{y_0\, \partial_j T_{0k}(y)
-y_j\, \partial_0 T_{0k}(y) + T_{jk}(y) - \delta_{jk} T_{00}(y) \Big\}\; ,
\ee
and the integrated WIs can thus be written as 
\bea\label{eq:boost1}
& & \int_{\partial R} d \sigma_\mu(x)\,\langle K_{\mu;0j}(x)\, T_{0k}(y)\, 
O_1 \dots O_n \rangle =\\ 
& & - \langle\Big\{y_0\partial_j T_{0k}(y) - y_j \partial_0 T_{0k}(y) 
+ T_{jk}(y) - \delta_{jk} T_{00}(y) \Big\}\, O_1 \dots O_n\rangle \; .\nonumber 
\eea
By using Eq.~(\ref{eq:WIstd2}), and for correlation functions for 
which boundary terms can be 
neglected when integrating by parts in $d^3 y$, such as connected 
correlations functions with operators $O_i$ localized in space,
the WIs read
\be\label{eq:KPcomm}
\int_{\partial R} d \sigma_\mu(x)\, \langle K_{\mu;0j}(x) 
{\overline T}_{0k}(y_0)\, O_1 \dots O_n \ranglec =
\delta_{jk}\, \langle {\overline T}_{00}(y_0)\, O_1 \dots O_n \ranglec \; .
\ee
Being valid for every string of external localized operators, 
i.e. for a generic state, this is the Euclidean version of the commutation 
relation of the momentum with the charges associated with the boosts. Finally 
it is interesting to notice that the number of fields $T_{\mu\nu}$ inserted 
in the correlation functions entering the two sides of these WIs is different.

\section{Momentum distribution from shifted boundaries\label{sec:SyCPI}}
In this	section	we show	how the	correlation functions of the total
momentum can be extracted in the Euclidean path integral formalism by
generalizing the periodic temporal boundary condition to shifted
boundary conditions. The relative contribution to the partition function 
of the states with momentum $\bp$ is ($L_0$ dependence suppressed)
\be\label{eq:SyCPI}
\frac{R({\bp})}{V} = 
\frac{{\rm Tr}\{e^{- L_0 \hat {\rm H}}\, \hat{\rm P}^{({\bp})}\}}{{\rm Tr}\{e^{- L_0 \hat {\rm H}}\}}\; , 
\ee
where the trace is over all the states of the Hilbert space, $\hat{\rm
  P}^{({\bp})}$ is the projector onto those states with total momentum
$\bp$, and $\hat {\rm H}$ is the Hamiltonian of the theory. If we 
introduce the partition function
\be
Z(\bz)=\Tr\{e^{- L_0 \hat {\rm H}}\, e^{i\hat\bp\bz}\}\; , 
\ee
in which states of momentum $\bp$ are weighted by a phase $e^{i\bp\cdot\bz}$, 
and we use the standard group theory machinery 
(see Ref.~\cite{DellaMorte:2010yp} for a detailed discussion on this point)
it easy to show that 
\be\label{eq:bella2}
R({\bp}) = \frac{1}{Z}\, \int d^3{\bz}\, e^{-i{\bp}\cdot{\bz}}\, Z(\bz)\; ,
\ee
where $Z=Z({\bf 0})$ is the ordinary thermal partition function.
The generating function $\Fgen({\bz})$ of the cumulants of  the momentum 
distribution is defined as usual as  
\be\label{eq:freeE}
e^{-\Fgen({\bz})} = 
\frac{1}{V} \sum_{\bp} e^{i{\bp}\cdot{\bz}} \, R({\bp})\;,
\ee
and the cumulants are given by 
\be\label{eq:cum1}
k^{_V}_{\{2 n_1, 2 n_2, 2 n_3\}} = (-1)^{n_1+n_2+n_3+1}\, 
\frac{\partial^{2n_1}}
{\partial \bz_1^{2n_1}} \frac{\partial^{2n_2}}{\partial \bz_2^{2n_2}}
\frac{\partial^{2n_3}}{\partial \bz_3^{2n_3}}
\frac{\Fgen({\bz})}{V}\Big|_{\bz=0}\; ,
\ee
where they have been normalized so as to have a finite limit when $V\rightarrow\infty$. 
The shifted partition function $Z(\bz)$ can be expressed as a Euclidean path 
integral with the field satisfying the boundary conditions
\be
\phi(L_0,{\bx})=\phi(0,{\bx}+{\bz})
\ee
in the compact direction. From Eqs.~(\ref{eq:bella2}) and  (\ref{eq:freeE}),
the generating function can thus be written as the ratio of partition functions
\be\label{eq:exp-F}
e^{-\Fgen({\bz})} = \frac{Z(\bz)}{Z}\;,
\ee 
i.e. two path integrals with the same action but different boundary 
conditions, and the cumulants can be obtained by 
deriving it with respect to the shift parameter $z$ an appropriate 
number of times. Since the cumulants are connected correlation functions
of the total momentum charge of the theory, they are ultraviolet 
finite as they stand (see also section \ref{sec:wi2} and appendix \ref{app:ren}). 
Therefore $\Fgen({\bz})$ and the momentum distribution $R({\bp})$ are finite as 
well, and they do not need any ultra-violet renormalization.

\subsection{Extension to the lattice}
When defined on a lattice, the theory is invariant under a discrete
subgroup of translations and rotations only, the momenta are quantized, 
and the continuum WIs are broken by discretization effects. Generic lattice 
definitions of the energy-momentum tensor, as well as the corresponding
charges, require ultraviolet renormalization. It is still possible, however, 
to factorize the Hilbert space of the lattice theory in sectors with definite 
conserved total momentum. The formula for the lattice projector 
$\hat{\rm P}^{({\bp})}$ can be easily obtained, and its explicit form can be found 
in Ref.~\cite{DellaMorte:2010yp}. Since only physical states contribute to the 
symmetry constrained path integrals in Eq.~(\ref{eq:SyCPI}),
the lattice momentum distribution $R({\bp})$ is expected to converge to the 
continuum universal one without the need for any ultra-violet renormalization.   
The definition of the cumulants in Eq.~(\ref{eq:cum1}) is thus applicable at 
finite lattice spacing, provided the derivatives are replaced with their discrete 
counterpart, and no additive or multiplicative ultraviolet-divergent 
renormalization is needed for taking the continuum limit.

\section{Two-point correlators of $T_{\mu\mu}$\label{sec:wi1}}
The two-point correlation functions of the diagonal components of 
the energy-momentum tensor satisfy WIs which are non-trivial and 
interesting in the thermal theory at finite volume. 
By choosing $T_{kk}(y)$ as interpolating operator, 
$\epsilon_\nu = \delta_{\nu 0} \epsilon_0$ and 
by using translational 
invariance and parity, the WI (\ref{eq:WIstd1}) can be written as 
\be\label{eq:00kk1}
\partial_0^x \Big\{\langle {\overline T}_{00}(x_0)\, T_{kk}(y)\rangle + 
\delta(x_0-y_0)\, \langle T_{00} + {\cal L}\rangle\Big\} = 0\; .
\ee
Analogously by taking $\epsilon_\nu = \delta_{\nu k} \epsilon_k$, 
$T_{00}(z)$ as interpolating operator, and 
by using translational invariance and ``$k$-parity'' 
\be
\partial_k^w \Big\{\langle {\widetilde T}_{kk}(w_k)\, T_{00}(z)\rangle + 
\delta(w_k-z_k)\, \langle T_{kk} + {\cal L}\rangle\Big\} = 0\; ,
\ee
where 
\be
{\widetilde T}_{\mu\nu}(w_k) = \int \Big[\prod_{\rho\neq k} d w_\rho\Big] \, T_{\mu\nu}(w)\; .
\ee
By integrating both of them and by taking the difference, we
obtain ($x_0 \neq y_0\; , w_k \neq z_k$)
\be\label{eq:00kk}
L_0\,\langle {\overline T}_{00}(x_0)\, T_{kk}(y)\rangle - 
L_k\,\langle {\widetilde T}_{kk}(w_k)\, T_{00}(z)\rangle = 
\langle T_{00}\rangle -\langle T_{kk}\rangle \; .     
\ee
If we subtract from both terms on the l.h.s the disconnected piece, we 
arrive at
\be\label{eq:belc}
L_0\, \langle {\overline T}_{00}(x_0)\, T_{kk}(y)\rangle_c -
L_k\, \langle \widetilde{T}_{kk}(w_k)\, T_{00}(z)\rangle_c
= \langle T_{00} \rangle - \langle T_{kk} \rangle \; . 
\ee
Non-trivial properties of Eqs.~(\ref{eq:00kk})-(\ref{eq:belc}) are that all operators 
are localized at non-zero physical distance from each other, that the number of 
fields $T_{\mu\mu}$ inserted in the correlation functions entering  the two sides of each 
equation is different, and that the additive renormalization counter-terms proportional 
to $a_1$ and $a_2$ in Eq.~(\ref{eq:Tren}) do not contribute. These equations are thus valid 
for the renormalized energy-momentum tensor as well, or conversely they imply that the 
overall renormalization constant satisfies ${\cal Z}=1$ (see appendix \ref{app:ren}). 
In regularizations that break translational invariance, they can be imposed to compute 
non-perturbatively the overall renormalization constant of the diagonal component 
of the field $T_{\mu\nu}$.

\subsection{Thermodynamic limit and finite-size effects}
In the thermodynamic limit the second term on 
the l.h.s. of Eq.~(\ref{eq:belc}) vanishes, and the infinite volume 
WI reads
\be\label{eq:beld}
L_0\, \langle {\overline T}_{00}(x_0)\, T_{kk}(y)\rangle_c 
= \langle T_{00} \rangle - \langle T_{kk} \rangle \; . 
\ee
The very same WI can be obtained in a slightly more elegant form.
By integrating a combination of two of the WIs in Eq.~(\ref{eq:WIstd1}), 
and by using translational invariance and parities we obtain
\be
\int d^4 x\, \langle x_0\, \partial_\mu T_{\mu 0}(x)\, T_{kk}(y) - 
                   x_k\, \partial_\mu T_{\mu k}(x)\, T_{00}(z) \rangle 
= \langle T_{00}\rangle - \langle T_{kk}\rangle\; ,  
\ee 
from which it is easy to derive Eq.~(\ref{eq:beld}). The latter
has a straightforward thermodynamic interpretation. If we remember that in 
the Euclidean
\be
\langle T_{00} \rangle = -e\; , \qquad \langle T_{kk} \rangle = p\; ,
\ee 
where $e$ and $p$ are the energy density and the pressure in the thermodynamic 
limit respectively, Eqs.~(\ref{eq:devL0O}) and (\ref{eq:beld}) imply the 
well-known thermodynamic relation 
\be\label{eq:termo2}
T \frac{\partial p}{\partial T}= e+p \; . 
\ee
By defining the entropy density of the finite-volume system as usual
\be
s^{_V} = \frac{1}{V}\frac{\partial}{\partial T}\Big[T\ln Z\Big]\; ,  
\ee
and by using Eq.~(\ref{eq:devL0O}) and the analogous one in the 
k-direction (see Eq.~\ref{eq:terk0})), the extensivity of the free 
energy and Eq.~(\ref{eq:termo2}) lead to
\be\label{eq:termo22}
Ts=e+p\qquad \Longrightarrow\qquad \frac{\partial}{\partial T}\; p = s\; .
\ee
The leading finite-size corrections in these thermodynamic potentials were 
calculated in Ref.~\cite{Meyer:2009kn} for a generic theory with a mass gap 
$M$ in the screening spectrum. For the sum of the energy and pressure 
they read
\be\label{eq:fsc1}
\< T_{kk}\> - \< T_{00} \>  - (e+p) = - \frac{\nu M T}{2\pi L} 
\Big[M + 3 T \frac{\partial M}{\partial T} \Big] e^{-ML} + \dots\; , 
\ee
where the factor $\nu$ stands for the multiplicity of the 
lightest screening state, and the dots stand for terms 
which vanish with a larger exponent. The WI~(\ref{eq:belc}) generalizes 
Eq.~(\ref{eq:termo2}) to finite-volume, and it shows that the violations in 
the latter are due to the finite-volume dependence of the energy density, i.e. 
\be\label{eq:FSE1}
\< T_{kk}\> - \< T_{00} \>  - T\frac{\partial \<T_{kk}\>}{\partial T}  
= L_k \frac{\partial \<T_{00}\>}{\partial L_k}\; .
\ee
By inserting the finite-size corrections computed in 
Ref.~\cite{Meyer:2009kn} we obtain 
 \be
T\, \frac{\partial \<T_{kk}\>}{\partial T} -\< T_{kk}\> + \< T_{00} \>   
= \frac{\nu M^2 T^2}{2\pi} \left(\frac{\partial M}{\partial T}\right) 
\left[ 1 + \frac{1}{M L}\right] e^{-M L} +\dots\; , 
\ee
and analogously for the violation of first equation (\ref{eq:termo22}) 
\be\label{eq:termo23}
T s^{_V} - \< T_{kk}\> + \< T_{00} \> = 
\frac{\nu T}{2\pi L^3} 
\Big[ 3 + 3 M L + M^2 L^2 \Big] e^{-M L} +\dots 
\ee
Finite-volume effects in the thermodynamic relations (\ref{eq:termo2})
and (\ref{eq:termo22}) are exponentially small in $ML$, and the leading
finite-size corrections are known functions of the lightest screening 
mass, its temperature derivative and the multiplicity of the associated 
state.  

\section{Two-point correlators of $T_{0k}$\label{sec:wi2}}
In a finite box a generic boost transformation rotates periodic 
fields to non-periodic ones. Those transformations are incompatible 
with the boundary conditions of the finite-volume theory. The 
integrated WIs associated with the $SO(4)$ rotations must be modified 
by finite-size contributions which vanishes in the thermodynamic limit. 
On the other hand the finite-volume theory has an energy-momentum tensor 
which is locally conserved, and the integrated WIs associated with a generic 
infinitesimal rotation $\omega_{\alpha\beta}$ can be constructed 
starting from Eqs.~(\ref{eq:WIstd1}) and by multiplying both sides 
by the proper factors. By choosing $\epsilon_\nu = \delta_{\nu k} \epsilon_k$,
$T_{0k}(y)$ as interpolating operator, and by using translational 
invariance and parity, the WI (\ref{eq:WIstd1}) can be written as 
\be\label{eq:0k0k1}
\partial_0^x \Big\{\langle {\overline T}_{0k}(x_0)\, T_{0k}(y)\rangle -
\delta(x_0-y_0)\, \langle T_{kk} + {\cal L}\rangle\Big\} = 0\; .
\ee
Analogously by taking $\epsilon_\nu = \delta_{\nu 0} \epsilon_0$,
$T_{0k}(z)$ as interpolating operator, and by using translational 
invariance and parity
\be
\partial_k^w \Big\{\langle {\widetilde T}_{0k}(w_k)\, T_{0k}(z)\rangle - 
\delta(w_k-z_k)\, \langle T_{00} + {\cal L}\rangle\Big\} = 0\; .
\ee
By integrating both of them and by taking the difference 
we obtain\footnote{This relation is explicitly verified for the 
free theory in appendix~\ref{sec:WIfree}.} 
($x_0 \neq y_0\; , w_k \neq z_k$)
\be\label{eq:bellissima}
L_0\, \langle {\overline T}_{0k}(x_0)\, T_{0k}(y) \rangle - L_k 
\langle \widetilde{T}_{0k}(w_k)\, T_{0k}(z) \rangle =
\langle T_{00} \rangle - \langle T_{kk} \rangle \; .
\ee
This WI can be derived directly from Eq.~(\ref{eq:boost1}) by paying attention
to the fact that boundary terms in the integration by parts cannot be neglected.
The comments below Eq.~(\ref{eq:beld}) apply also in this case. In particular, 
in regularizations that break translational invariance, these equations can be 
imposed to compute non-perturbatively the overall renormalization constant of 
the off-diagonal components of the field $T_{\mu\nu}$. These WIs are reminiscent
of those associated with non-singlet chiral symmetry in  QCD,  which turn out to be so 
relevant to renormalize the axial current when chiral symmetry is explicitly 
broken by the regularization~\cite{Bochicchio:1985xa,Luscher:1996jn} (see also 
Ref.~\cite{Luscher:1998pe} for a review). From Eqs.~(\ref{eq:belc}) and 
(\ref{eq:bellissima}) we obtain
\be
\hspace{-0.015cm}\langle {\overline T}_{0k}(x_0) T_{0k}(y) \rangle -
\langle {\overline T}_{00}(x'_0) T_{kk}(y')\rangle_c =
\frac{L_k}{L_0}\Big\{\langle \widetilde{T}_{0k}(w_k) T_{0k}(z) \rangle -
\langle \widetilde{T}_{kk}(w'_k) T_{00}(z')\rangle_c\Big\}\, ,
\label{eq:bell3}
\ee
where in all correlators the operators must be inserted at 
a physical distance ($x_0\neq y_0$, etc.). It is interesting to notice that this 
relation can be derived directly from the conservation of the energy-momentum 
tensor in Eq.~(\ref{eq:WIstd2}) without explicit reference to the transformation
properties of the interpolating operators, see Ref.~\cite{Giusti:2010bb} and 
below.

\subsection{Thermodynamic limit and finite-size effects}
In the infinite-volume limit the second term on the l.h.s. of 
Eq.~(\ref{eq:bellissima}) vanishes, and the relation reads
\be\label{eq:bella}
L_0\, \langle {\overline T}_{0k}(x_0)\, T_{0k}(y) \rangle = \langle T_{00}\rangle - 
\langle T_{kk}\rangle\; .
\ee
Its thermodynamic interpretation is straightforward. If we remember that in the 
Euclidean the momentum operator maps to  
$\hat p_k \rightarrow -i {\overline T}_{0k}$, then ($x_0 \neq y_0$) 
\be\label{eq:bho1}
\langle {\overline T}_{03}(x_0)\, T_{03}(y) \rangle = - k_{\{0,0,2\}}\; .
\ee
The Eq.~(\ref{eq:bella}) can then be written as 
\be\label{eq:bho2}
k_{\{0,0,2\}} = T (e+p) = T^2 s\; .
\ee
The WI~(\ref{eq:bellissima}) generalizes this equation to finite-volume.
The second term on the left-hand side of (\ref{eq:bellissima}) vanishes 
exponentially in the lowest 
screening level corresponding to a state with a non-zero momentum in the 
time direction. Whenever the lightest screening state 
has vanishing momentum in the time direction, which is expected to be 
the (generic) case, the leading finite-volume corrections to the 
second cumulant are those from the r.h.s. of the Eq.~(\ref{eq:bellissima}). 
They are exponentially small in $ML$, and their explicit form is given 
by the r.h.s. of Eq.~(\ref{eq:fsc1}) multiplied by $T$.

\section{Recursion relation for $2n$-point correlators of 
${\overline T}_{0k}$\label{sec:wi3}}
In the thermodynamic limit the WIs (\ref{eq:WIstd1}) imply a recursion 
relation among correlators of ${\overline T}_{0k}$ inserted at a physical 
distance. It can be derived by repeatedly using Eq.~(\ref{eq:WIstd1}) 
with different strings of interpolating operators. This can be concisely 
shown by introducing a new action
\be\label{sec:Sj}
S_J = S -\int d^4 x\, J(x_0)\, T_{0k}(x)
\ee
which differs from the standard one by a term which 
couples an external source $J(x_0)$, constant over
the time-slices, with the momentum field in direction 
$k$. The corresponding path-integral is defined as 
\be
Z[J] = \int D \phi\; e^{-S_J}\; ,  
\ee
and the expectation value of a generic operator 
$\langle O \rangle_J$ is defined as usual. It turns 
out to be useful to introduce the operators ${\cal D}_{[x_0^1..x_0^n]}$
defined as 
\be\label{eq:Dxx}\displaystyle 
{\cal D}_{[x_0^1..x_0^n]}\, F(J) =
\frac{\partial}{\partial J(x_0^n)}\, \dots\, \frac{\partial}{\partial J(x_0^1)}\,
F(J)\, {\Big|}_{J=0}\; , 
\ee
where $F(J)$ is a generic functional of the external source and the $x_0^i$ are all different. 
As usual when applied to $F(J)=\ln{Z[J]}$ it gives the {\it connected} correlation functions
of $n$ momentum operators ${\overline T}_{0k}(x_0^i)$ inserted at a physical distance. 
By choosing $\epsilon_\nu = \delta_{\nu k} \epsilon_k$ (no summation over $k$), 
$\widetilde{\overline{T}}_{00}$ as interpolating operator, where
\be
\widetilde{\overline{T}}_{\mu\nu}(x) = \int \Big[\prod_{\rho\neq 0,k} d x_\rho\Big]\, T_{\mu\nu}(x)\; ,
\ee
and 
by using translational invariance in time\footnote{When an operator $D_{[x_0^1..x_0^{2n}]}$
is applied, the action entering the definition of the correlation functions 
is the standard one.}, the WI (\ref{eq:WIstd1}) gives
\be\label{eq:00kkJ12}
\partial^{x^1}_0 D_{[x_0^3..x_0^{2n}]} \langle \widetilde{\overline{T}}_{00}(x^1)\, 
\widetilde{\overline{T}}_{0k}(x^2)\rangle_{J,c} =
\partial^{x^2}_k D_{[x_0^3..x_0^{2n}]} \langle \widetilde{\overline{T}}_{kk}(x^2)\, 
\widetilde{\overline{T}}_{00}(x^1)\rangle_{J,c}\; .
\ee
Analogously by taking $\epsilon_\nu = \delta_{\nu 0} \epsilon_0$, 
$\widetilde{\overline{T}}_{0k}$ as interpolating operator, and thanks to 
translational invariance and the symmetry of $T_{0k}$, the WI (\ref{eq:WIstd1}) leads also to 
\be\label{eq:00kkJ2}
\partial^{x^1}_0 D_{[x_0^3..x_0^{2n}]} \langle \widetilde{\overline{T}}_{00}(x^1)\, 
\widetilde{\overline{T}}_{0k}(x^2)\rangle_{J,c} =
\partial^{x^2}_k D_{[x_0^3..x_0^{2n}]} \langle \widetilde{\overline{T}}_{0k}(x^1)\, 
\widetilde{\overline{T}}_{0k}(x^2)\rangle_{J,c}\; .
\ee
By putting together last two WIs we arrive at 
\be
\partial^{x^2}_k D_{[x_0^3..x_0^{2n}]}
\Big\{\langle \widetilde{\overline{T}}_{0k}(x^2)\, 
\widetilde{\overline{T}}_{0k}(x^1)\rangle_{J,c} -
\langle \widetilde{\overline{T}}_{kk}(x^2)\, 
\widetilde{\overline{T}}_{00}(x^1)\rangle_{J,c}\Big\} = 0\; . 
\ee
Since the argument of the partial derivative is constant in $x^2_k$, we can integrate
by keeping all insertions at a physical distance ($x_0^i$ all different) and obtain 
\bea
& & D_{[x_0^3..x_0^{2n}]} \Big\{ 
\langle {\overline T}_{0k}(x_0^1)\, {\overline T}_{0k}(x_0^2) \rangle_{J,c} - 
\langle {\overline T}_{00}(x_0^1)\,  {\overline T}_{kk}(x_0^2)  \rangle_{J,c} \Big\} 
=\nonumber\\[0.25cm] 
& & L_k^2\; D_{[x_0^3..x_0^{2n}]} 
\Big\{\langle \widetilde{\overline{T}}_{0k}(x^1)\, 
\widetilde{\overline{T}}_{0k}(x^2) \rangle_{J,c} - 
\langle \widetilde{\overline{T}}_{00}(x^1)   \widetilde{\overline{T}}_{kk}(x^2)\, 
\rangle_{J,c} \Big\} \; , 
\eea
which generalizes Eq~(\ref{eq:bell3}). The Eqs.~(\ref{eq:terk}) and 
(\ref{eq:appbbel}) give 
\be\label{eq:blla}
{\cal D}_{[x_0^3..x_0^{2n}]}\,\Big\{ L_0\, \langle {\overline T}_{kk}(x^2_0) \rangle_J - 
L_k\, \frac{\partial}{\partial L_k} \ln{Z[J]}\, - \int d w_0\, J(w_0)\, 
\langle {\overline T}_{0k}(w_0) \rangle_J\; , 
\Big\} = 0\; 
\ee
where again the $x_0^i$ are all different, and (a generalization of) 
Eq.~(\ref{eq:devL0O}) leads to  
\be
D_{[x_0^3..x_0^{2n}]} \langle {\overline T}_{00}(x_0^1)\, {\overline T}_{kk}(x_0^2)  \rangle_{J,c}
= \frac{\partial}{\partial L_0}\; D_{[x_0^3..x_0^{2n}]} \langle {\overline T}_{kk}(x_0^2) 
\rangle_J\; .
\ee
Finally, by putting together the last three equations, we can write (all 
insertions at a physical distance from each other)
\bea
& & \langle {\overline T}_{0k}(x_0^1) 
\dots {\overline T}_{0k}(x_0^{2n}) \rangle_c =  
(2 n -1)\, \frac{\partial}{\partial L_0}\,\Big\{\frac{1}{L_0}\,
\langle {\overline T}_{0k}(x_0^3) 
\dots {\overline T}_{0k}(x_0^{2n}) \rangle_c \Big\} + \nonumber\\[0.25cm]
& & \hspace{-0.5cm} L_k^2 
\Big\{\langle \widetilde{\overline{T}}_{0k}(x^1)\, 
\widetilde{\overline{T}}_{0k}(x^2) {\overline T}_{0k}(x_0^3)\dots 
{\overline T}_{0k}(x_0^{2n}) \rangle_c - 
\langle \widetilde{\overline{T}}_{00}(x^1)\, \widetilde{\overline{T}}_{kk}(x^2)\, 
{\overline T}_{0k}(x_0^3)\dots {\overline T}_{0k}(x_0^{2n})\rangle_c \Big\}\nonumber\\[0.25cm]
& & + \frac{\partial}{\partial L_0}
\Big\{\frac{1}{L_0}\Big[L_k\frac{\partial}{\partial L_k}-1\Big] \langle {\overline T}_{0k}(x_0^3) 
\dots {\overline T}_{0k}(x_0^{2n}) \rangle_c \Big\}\; .\label{eq:bigg} 
\eea

\subsection{Thermodynamic limit and finite-size effects}
The last three terms in Eq.~(\ref{eq:bigg}) are finite-size effects which 
vanish in the infinite volume limit: the first two because the distance 
$|x_k^1-x_k^2|$ can be arbitrarily large, the third one due to the expected 
volume dependence of the correlation function. In the thermodynamic limit 
we thus arrive at the recursive relation  
\be\label{eq:recKK}
\langle {\overline T}_{0k}(x_0^1) 
\dots {\overline T}_{0k}(x_0^{2n}) \rangle_c =  
(2 n -1)\, \frac{\partial}{\partial L_0}\,\Big\{\frac{1}{L_0}\,
\langle {\overline T}_{0k}(x_0^3) 
\dots {\overline T}_{0k}(x_0^{2n}) \rangle_c \Big\}\; ,
\ee
with a straightforward thermodynamic interpretation
\be\label{eq:recK}
k_{\{0,0,2n\}} = (2n-1) T^2 \frac{\partial}{\partial T}\Big\{T\, k_{\{0,0,2n-2\}} \Big\}\; .
\ee
Cumulants with non-trivial indices in the other two spatial 
directions are related to those in Eq.~(\ref{eq:recK}) by cubic 
symmetry. For the total momentum, Eq.~(\ref{eq:recKK}) is the analog
of the well-known recursive relation among the connected correlation
functions of the energy, see Eq.~(\ref{eq:devL0O}) with ${\cal O}$ 
being a string of $\overline{T}_{00}$'s. The Eq.~(\ref{eq:recK}) is 
checked explicitly in appendix~\ref{app:freeZ} for the free theory. 
It is interesting to notice that it leads to a straightforward 
physical interpretation of the fourth cumulant, i.e. 
\be\label{eq:Kcv}
k_{\{0,0,4\}} = 3T^4 c_v + 9T^3(e+p)\; , 
\ee
where $c_v$ is the specific heat of the system in the thermodynamic limit.
Finite-volume corrections to this identity can be computed 
by starting from the WIs (\ref{eq:WIstd1J}) and by following 
a procedure analogous to the one which lead to Eq.~(\ref{eq:bellissima}).
For the fourth cumulant the relevant WI reads\footnote{In this and next equation 
the arguments $x_0^i$ and $x_k^i$ are omitted. Operators are inserted always at 
a physical distance.}
\bea  \label{eq:bellissima4}\displaystyle
&& L_0^3\,  \< {\overline T}_{0k} {\overline T}_{0k} {\overline T}_{0k} T_{0k}\>_c
- L_k^3\, \< {\widetilde T}_{0k} {\widetilde T}_{0k}
{\widetilde T}_{0k} T_{0k}\>_c 
=3\, \Big\{\<T_{00}\>  - \<T_{kk}\>\Big\}   
\\[0.25cm] 
&& -3\, \Big\{L_0\, \<{\overline T}_{00} T_{00}\>_c  - L_k\, \<{\widetilde T}_{kk} T_{kk}\>_c \Big\}
+6\,  \Big\{L^2_0\, \<{\overline T}_{0k} {\overline T}_{0k} T_{00}\>_c  - L^2_k\, 
\<{\widetilde T}_{0k} {\widetilde T}_{0k} T_{kk}\>_c \Big\}\; .  
\nonumber
\eea
To compute the leading finite-size effects, it is convenient to rewrite 
this equation as 
\bea
&& L_0^3\,  \< {\overline T}_{0k} {\overline T}_{0k} {\overline T}_{0k} T_{0k}\>_c = 
3\, \Big\{L_0 \frac{\partial}{\partial L_0} - L_k \frac{\partial}{\partial L_k} -2\Big\}
\Big\{\<T_{00}\> - \<T_{kk}\>\Big\}\nonumber\\[0.25cm]
&& + L_k^3\, \< {\widetilde T}_{0k} {\widetilde T}_{0k} {\widetilde T}_{0k} T_{0k}\>_c 
- 6 L_k\Big\{1 + L_k \frac{\partial}{\partial L_k} - L_0 \frac{\partial}{\partial L_0} \Big\}
\,\<{\widetilde T}_{0k} T_{0k}\>\; .
\eea
The second line of the right-hand side vanishes exponentially in the lowest
screening level corresponding to a state with non-zero momentum in the time 
direction. We therefore turn our attention to the first line, from which we 
compute the leading finite-size corrections to the fourth cumulant 
by using again the results in Ref.~\cite{Meyer:2009kn} 
\bea
\hspace{-0.375cm}&& 
\displaystyle T^{-3}\,  \< {\overline T}_{0k} {\overline T}_{0k} {\overline T}_{0k} T_{0k}\>_c - 
3 T c_v - 9 (e+p) = 
- \frac{3 \nu T}{2\pi L}\Big\{ 3\, \frac{\partial }{\partial T} 
\Big(e^{-M L} M T^2\, {\textstyle\frac{\partial M}{\partial T}}\Big) +\nonumber\\[0.25cm] 
\hspace{-0.375cm}&&
\displaystyle e^{-M L}\, M T\, \frac{\partial M}{\partial T}\, (7 - 2 M L)
-\frac{e^{-M L}}{L^2}\, (6 + 6ML + M^3L^3 ) 
\Big\}
+\dots
\eea
where the dots stand for terms that vanish exponentially faster.

\section{Main results and conclusions\label{sec:mainres}}
By comparing the WI~(\ref{eq:bho2}) and the Eq.~(\ref{eq:cum1}) with $n_3=1$ and
$n_1=n_2=0$, the entropy density in the thermodynamic 
limit can be written as 
\be\label{eq:final1}
s = - \frac{1}{T^2}\,\lim_{V\rightarrow \infty} \frac{1}{V} \frac{d^2}{d z^2}
\ln{Z(\{0,0,z\})} \Big|_{z=0}\; .
\ee
Thanks to Eq.~(\ref{eq:termo22}), the pressure can be 
computed by integrating $s$ in the temperature,
and the ambiguity left due to the integration constant is consistent 
with the fact that $p$ is defined up to an arbitrary additive 
renormalization constant. From the relation (\ref{eq:Kcv}), 
the specific heat of the system can be written as 
\be\label{eq:final2}
c_v =  \lim_{V\rightarrow \infty} \frac{1}{V}\left[
\frac{1}{3 T^4}\frac{d^4}{d z^4} + \frac{3}{T^2} \frac{d^2}{d z^2}\right]
\ln{Z(\{0,0,z\})} \Big|_{z=0}\; .
\ee
The last two equations make clear that the response of the partition 
function to the shift $z$ is governed by basic thermodynamic properties 
of the system, and that the potentials entering the equation of state of 
the thermal theory can be extracted by rather simple formulas. It is worth 
stressing that Eqs.~(\ref{eq:final1}) and (\ref{eq:final2}) are the result of a 
judicious combination of the WIs associated with invariance of the theory under 
the Poincar\'e group only. The relevant WIs relate correlation functions of conserved 
charges, and this is why they are free from ultraviolet subtractions and 
renormalizations. The formulas (\ref{eq:final1}) 
and (\ref{eq:final2}) are indeed valid for a wider class of thermal 
field theories than the scalar one~\cite{Giusti:2010bb}. For gauge theories, 
however, their derivation involves additional complications due to the 
non-commutation of translations and gauge transformations, and it will be 
presented in a forthcoming paper~\cite{Giusti:2011}.

Crucially the Eqs.~(\ref{eq:final1}) and (\ref{eq:final2}) remain 
valid in a lattice box, up to exponentially suppressed finite-size effects and 
harmless discretization errors, provided the derivatives are replaced with their 
discrete counterpart. In a finite volume  some of the relevant WIs generalize 
well-known thermodynamic relations, and they are the basic ingredient to relate 
the finite-size effects in the cumulants to those in simple quantities such as 
the energy density and pressure which are known~\cite{Meyer:2009kn}. 
With respect to the analogous observables computed 
in the standard methods~\cite{Engels:1981qx,Engels:1990vr,Endrodi:2007tq,Meyer:2009tq}, 
the entropy density and the specific heat computed from the equations above do require 
neither a vacuum subtraction nor an ultraviolet renormalization constant to be fixed. 
Moreover an improvement of the action automatically 
leads to a corresponding improvement in the thermodynamic quantities. 
Conversely the shifted boundary conditions adopted in 
this paper and the WIs (\ref{eq:belc}) and (\ref{eq:bellissima}), which are valid 
for the renormalized energy-momentum tensor as well, can be combined to design a 
non-perturbative renormalization procedure for $T_{\mu\nu}$ on the lattice.

The generating function $\Fgen({\bz})$ may turn out to be an interesting
thermodynamic quantity in itself. For a scale-invariant theory, for instance,
one might have expected that $\Fgen({\bz})$ can be an arbitrary
function of $T|\bz|$.
However, the recurrence relation (\ref{eq:recK}) and the scale invariance
fix its functional form unambiguously to be
\be\label{eq:final}
\frac{\Fgen({\bz})}{V} =
\frac{s}{4} \left[1- \frac{1}{(1+T^2 \bz^2)^2}\right]\, ,
\ee
no matter what the coupling of the theory is. This is particularly relevant to
those strongly coupled theories that can be treated with the AdS/CFT correspondence. 
We note that for $\bz$ purely imaginary, ${\bf v} \equiv iT\bz$ can be interpreted 
as the macroscopic velocity of the thermal system. The double poles in $\bz$ 
appearing in Eq.~(\ref{eq:final}) corresponding to $|{\bf v}|=1$ therefore do not 
come as a surprise.

\section*{Acknowledgments}
We thank Michele Della Morte and Martin L\"uscher for useful
discussions. HBM's work is supported by the \emph{Center for Computational Sciences} 
in Mainz.
\appendix 

\section{Renormalization pattern of $T_{\mu\nu}$\label{app:ren}}
The renormalization pattern of the energy-momentum tensor was 
studied in great detail in perturbation 
theory~\cite{Callan:1970ze,Collins:1976vm,Brown:1979pq,Hathrell:1981zb,Caracciolo:1988hc}.
Here we assume to work in a regularization which 
preserves translational invariance, such as the dimensional 
one~\cite{Callan:1970ze,Collins:1976vm,Brown:1979pq,Hathrell:1981zb}.
A more general analysis, which is needed for the lattice theory 
can be found in~\cite{Caracciolo:1988hc}. Since the $\phi^4$ theory 
is not asymptotically free, it is difficult to extend the results reviewed  
here non-perturbatively. For the scope of this appendix, however, 
the standard power counting argument is sufficient.

The bare energy-momentum tensor is a dimension-4 field, symmetric under the 
exchange $\mu\leftrightarrow\nu$, even under the transformation $\phi \rightarrow -\phi$, 
and which satisfies the conservation equation (\ref{eq:WIstd2}). The renormalized
tensor can thus be written as   
\be
\widehat T_{\mu\nu} = {\cal Z}\left\{T_{\mu\nu} + a_1 
\left[\delta_{\mu\nu}\Box - \partial_\mu\partial_\nu \right]\phi^2 + a_2\, \delta_{\mu\nu}\right\}\; .
\ee 
Since by construction the last two fields on the r.h.s. have zero divergence, 
$\widehat T_{\mu\nu}$ satisfies the conservation equations (\ref{eq:WIstd2}) as well. 
If we choose the $O_i$ to be a combination of elementary fields, 
the integrated WIs (\ref{eq:mixWI}) can be written as 
\be\label{eq:renorm}
\frac{\epsilon_\nu}{\cal Z} \int_{\partial R} d \sigma_\mu(x)\, 
\langle \widehat T_{\mu\nu}(x) \widehat O_1 \dots \widehat O_n \rangle = -  
\sum_{i=1}^{n}\, \left\langle \widehat O_1\dots \delta_\epsilon \widehat O_i\dots 
\widehat O_n \right\rangle\; , 
\ee
where $\widehat O_i$ are the corresponding renormalized fields, and 
the contributions proportional to $a_1$ and $a_2$ vanish because they are surface 
integrals of vectors with null divergence.
Since the correlation functions entering Eq.~(\ref{eq:renorm}) are finite by construction, 
then also ${\cal Z}$ must be finite. If we choose ${\cal Z}=1$, the renormalized energy-momentum 
tensor reads
\be\label{eq:Tren}
\widehat T_{\mu\nu} = T_{\mu\nu} + a_1 
\left[\delta_{\mu\nu}\Box - \partial_\mu\partial_\nu \right]\phi^2 
+ a_2\, \delta_{\mu\nu}
\ee
and it satisfies WIs of the same form as those of the bare field, i.e. 
Eqs. (\ref{eq:WIstd2}) and (\ref{eq:mixWI}). The last two fields 
on the r.h.s. of Eq.~(\ref{eq:Tren}) do not contribute
to the charges ${\overline T}_{0k}(x_0)$, since the integration by parts of the 
term proportional to $a_1$ gives no boundary contributions.
The physical momentum fields are therefore defined 
by their bare expressions, and no renormalization ambiguity is left. 
The field proportional to $a_2$ contributes to the charge ${\overline T}_{00}(x_0)$
by an additive constant term $a_2 L^3$, which is ultraviolet divergent and 
proportional to the volume of the system. It contributes to the vacuum expectation value of the 
energy, but it cancels in connected correlation functions of ${\overline T}_{00}$ with other fields.  
An analogous analysis applies for the fields $\widetilde T_{k\nu}(x_k)$ ($\nu \neq k$) 
and $\widetilde T_{kk}(x_k)$. They are the momenta and the energy if direction 
$k$ is interpreted as the ``time'' direction. In particular the divergent coefficient 
$a_2$ is the very same for ${\overline T}_{00}(x_0)$ and $\widetilde T_{kk}(x_k)$.

The second term on the r.h.s. of Eq.~(\ref{eq:Tren}) (as well as the third one) 
contributes to the trace $T_{\mu\mu}$ of the energy momentum-tensor. In the quantum 
theory the renormalization procedure indeed breaks the conformal invariance 
of the classical theory also in the massless limit (trace anomaly).

\section{Ward identities in presence of external sources\label{sec:appSj}}
In this appendix we generalize some of the results of section \ref{sec:prel}
to the theory defined by the action $S_J$ given in Eq.~(\ref{sec:Sj}). The starting 
point are the the WIs  
\bea\label{eq:WIstd1J}
& & \epsilon_\nu\, \langle \Big\{ 
\partial_\mu T_{\mu\nu}(x) + J(x_0)\partial_\nu T_{0k}(x)
- \partial_0\left[J(x_0)(T_{\nu k}(x) + \delta_{\nu k} {\cal L}(x))\right]\\
& & - J(x_0) \partial_k [T_{\nu 0}(x) + \delta_{\nu 0}{\cal L}(x)] \Big\}\, 
O_1 \dots O_n \rangle_J = -  \sum_{i=1}^{n}\, \left\langle O_1 \dots 
\delta^x_\epsilon O_i \dots O_n \right\rangle_J\; ,  \nonumber
\eea
which extend those in Eq.~(\ref{eq:WIstd1}). 

\subsection{Evolution and translational invariance in $k$-direction}
From the previous WIs we can derive the ``k-time'' evolution of a 
generic field. If we choose $\epsilon_\nu = \delta_{\nu k} \epsilon_k$ 
(no summation over $k$) in Eq.~(\ref{eq:WIstd1J}), and we integrate
over a thick $k$-slice $R$ with the field $O_1$ being inserted into it 
while the operators $O_2\dots O_n$ are localized outside, we obtain
\be
\partial_k \, \langle  O_1(x^1)\, O_2\dots O_n\rangle_J
= - \int_{\partial R} d \sigma_k(x)\, \langle T_{kk}(x)\, O_1(x^1)\, O_2\dots O_n\ \rangle_J\; .
\ee
This is the Euclidean version of the $k$-time evolution of the generic field
$O_1(x^1)$, and therefore the field ${\widetilde T}_{kk}$  is the Hamiltonian 
of the system associated with the $k$-time. If $R$ covers the full space, we obtain
\be
\sum_{i=1}^{n}\, \partial^{x^i}_k \left\langle O_1(x^1) \dots 
O_i(x^i) \dots O_n(x^n) \right\rangle_J = 0\; ,
\ee 
i.e. a generic correlation function is translational invariant in the 
$k$ direction even in presence of the external source. This is expected 
since the source field is constant in $x_k$, and it does not break 
translational invariance in this direction. By following an analogous procedure, it is 
straightforward to show that translational invariance is preserved in 
the other two spatial directions too. Since the field ${\widetilde T}_{kk}$ is the 
Hamiltonian in direction $k$, we can write 
\be\label{eq:terk0}
\langle \widetilde{T}_{kk}(x_k)\rangle_J = \frac{\partial}{\partial L_k} \ln{Z[J]}\; ,
\ee
and the independence of the r.h.s from $x_k$ (translational invariance) implies  
\be\label{eq:terk}
\int d^4 x\, \langle T_{kk}(x)\rangle_J = 
L_k\, \frac{\partial}{\partial L_k} \ln{Z[J]}\; .
\ee 

\subsection{Translational invariance in 0-direction}
If we consider Eq.~(\ref{eq:WIstd1J}) for the interpolating 
field ${\overline T}_{kk}(y_0)$, we choose (no summation over $k$) 
$\epsilon_\nu = \epsilon_k \delta_{\nu k}$, 
and we integrate over the time-slices we obtain
\bea
\partial^x_0\, \langle {\overline T}_{0k}(x_0)\, {\overline T}_{kk}(y_0) \rangle_J & = &
\partial^x_0 \Big\{J(x_0)\, \langle \Big[{\overline T}_{kk}(x_0) + 
{\overline {\cal L}}(x_0)\Big]\, {\overline T}_{kk}(y_0)\rangle_J\Big\}\nonumber\\[0.25cm] 
& + & \Big[\partial^y_0 \delta(y_0-x_0)\Big] \langle {\overline T}_{0k}(y_0) \rangle_J\; . 
\label{eq:j1}
\eea
By taking again Eq.~(\ref{eq:WIstd1J}) for the field 
${\overline T}_{kk}(y_0)$ but for 
$\epsilon_\nu  = \epsilon_0 \delta_{\nu 0}$, 
and by integrating 
over the full space we arrive at
\be\label{eq:j2}
\partial^y_0\, \langle {\overline T}_{kk}(y_0) \rangle_J = -
\int d x_0\, J(x_0)\, \partial^x_0 \langle 
{\overline T}_{0k}(x_0)\, {\overline T}_{kk}(y_0)\rangle_J \; .
\ee
If we now put together Eqs.~(\ref{eq:j1}) and (\ref{eq:j2}) we 
obtain
\bea
\partial^y_0\, \langle {\overline T}_{kk}(y_0) \rangle_J & = & -
\int d x_0\, J(x_0)\, \partial^x_0 \Big[J(x_0) 
\langle ({\overline T}_{kk}(x_0) + {\overline {\cal L}}(x_0)) 
{\overline T}_{kk}(y_0)\rangle_J\Big]\nonumber\\[0.25cm] 
& - &\label{eq:apbb}
\Big[\partial^y_0 J(y_0)\Big] \langle {\overline T}_{0k}(y_0) \rangle_J\; . 
\eea
We are interested in applying the operator 
${\cal D}_{[x_0^1..x_0^{2n}]}$ defined in Eq.~(\ref{eq:Dxx}) 
on both sides of the last equation. Since the $x_0^i$ 
are all different from each other, the first term on the 
r.h.s. of Eq.~(\ref{eq:apbb}) does not contribute, and we obtain  
\be
{\cal D}_{[x_0^1..x_0^{2n}]}\,\Big\{
\partial^y_0\, \langle {\overline T}_{kk}(y_0) \rangle_J 
+ [\partial^y_0 J(y_0)]\, \langle {\overline T}_{0k}(y_0) \rangle_J
\Big\}
= 0\; ,
\ee
where at most one of the $x_0^i$ can coincides with $y_0$. In this case
\be
{\cal D}_{[x_0^1..x_0^{2n}]}\,\Big\{ J(y_0) \partial^y_0 \langle {\overline T}_{0k}(y_0) \rangle_J
\Big\} = 0
\ee
and, since the derivative $\partial^y_0$ commute with ${\cal D}_{[x_0^1..x_0^{2n}]}$,  
\be\label{eq:appbbel}
\partial^y_0\, {\cal D}_{[x_0^1..x_0^{2n}]}\,\Big\{
\langle {\overline T}_{kk}(y_0) \rangle_J + J(y_0) \, \langle {\overline T}_{0k}(y_0) \rangle_J
\Big\}
= 0\; .
\ee
It is reassuring to notice that this equation can be obtained 
directly from Eq.~(\ref{eq:WIstd1}) without the need of 
introducing $S_J$.\\

By putting together Eqs.~(\ref{eq:terk}) and (\ref{eq:appbbel}) 
we finally obtain Eq.~(\ref{eq:blla}), i.e. 
\be
{\cal D}_{[x_0^1..x_0^{2n}]}\,\Big\{ L_0\, \langle {\overline T}_{kk}(y_0) \rangle_J - 
L_k\, \frac{\partial}{\partial L_k} \ln{Z[J]}\, - \int d w_0\, J(w_0)\, 
\langle {\overline T}_{0k}(w_0) \rangle_J
\Big\} = 0\; ,
\ee
where $y_0$ in this equation is different from all $x_0^i$. 

\section{The Ward Identity (\ref{eq:bellissima}) in the free theory\label{sec:WIfree}}
In this appendix we check explicitly Eq.~(\ref{eq:bellissima}) 
for the free theory. Other WIs, for instance  Eqs.~(\ref{eq:belc}) 
and (\ref{eq:bell3}), can be verified analogously. In a finite $4$-dimensional 
volume the propagator
\be
\langle \phi(x) \phi(y) \rangle = D(x-y)
\ee
is given by
\be\label{eq:sump}\displaystyle
D(x) \equiv \frac{1}{V} \sum_p \frac{e^{ipx}}{p^2 + m^2} \; ,
\ee 
where $p$ runs over a $4$-dimensional momentum-space lattice 
$p_\mu= 2\pi\, n_\mu/L_\mu$ with $\mu=1,\dots,4$, and 
$n_\mu\in {\cal Z}$. It satisfies the equation of motion 
\be\label{eq:motionF}
\left\{\Box - m^2 \right\}D(x-y) = - \delta^{({\rm p})}(x-y)
\ee
where the periodic function $\delta^{({\rm p})}$ is defined as 
\be
\delta^{({\rm p})}(x) = \frac{1}{V} \sum_p e^{ipx} \; .
\ee
Thanks to the Wick theorem and to the invariance under parity 
\bea
& & \langle T_{\mu\nu}(x) T_{0 k}(y) \rangle =  \left[\partial_\mu\partial_0 D(x-y)\right]
                                        \left[\partial_\nu\partial_k D(x-y)\right] + 
                                        \left[\partial_\mu\partial_k D(x-y)\right] 
                                        \left[\partial_\nu\partial_0 D(x-y)\right]
\nonumber\\[0.25cm]  
& &\qquad  - \delta_{\mu\nu}\Big\{\left[\partial_\rho\partial_0 D(x-y)\right] 
                         \left[\partial_\rho\partial_k D(x-y)\right]   + 
                    m^2 \left[\partial_0 D(x-y)\right] \left[\partial_k D(x-y)\right] \Big\}\; ,
\eea
where the derivatives are with respect to $x$.
By using the equation of motion (\ref{eq:motionF}), it is easy to show that 
\be
\partial_\mu \langle T_{\mu\nu}(x) T_{0 k}(y) \rangle = 
-\left[\partial_0 \delta^{({\rm p})}(x-y)\right] \Big[\partial_\nu\partial_k D(x-y)\Big]
-\left[\partial_k \delta^{({\rm p})}(x-y)\right] \Big[\partial_\nu\partial_0 D(x-y)\Big]\; , 
\nonumber
\ee
and by remembering that 
\be
K_{\mu;0k}(x) = x_0 T_{\mu k}(x) - x_k T_{\mu 0}(x), 
\ee
the symmetry of $T_{\mu\nu}$ implies ($y_0\neq 0$ and $y_k\neq 0$, no summation over $k$)
\be\label{eq:eq:gg}
\int d^4 x\, \partial_\mu \langle K_{\mu;0k}(x) T_{0k}(y) \rangle = 
\partial_k^2 D(z)\Big|_{z=0} - \partial_0^2 D(z)\Big|_{z=0} \; .
\ee
By writing the r.h.s. as 
\be
\langle T_{00}\rangle - \langle T_{kk}\rangle = \partial_k^2 D(z)\Big|_{z=0} - \partial_0^2 D(z)\Big|_{z=0}
\ee
then Eq.~(\ref{eq:eq:gg}) corresponds to Eq.~(\ref{eq:bellissima}) in the free theory, i.e. 
\be
\int d^4 x\, \partial_\mu \langle K_{\mu;0k}(x) T_{0k}(y) \rangle = 
\langle T_{00}\rangle - \langle T_{kk}\rangle\; , \qquad
y_0 \neq0\;, \;\; y_k\neq 0\; . 
\ee

\section{Generating function for free and scale-invariant cases\label{app:freeZ}}
In this appendix we compute the generating function $\Fgen({\bz})$ of the 
cumulants of the momentum distribution in the free theory, and check explicitly the 
recursion relation (\ref{eq:recK}). We also compute the same quantity on the 
lattice to assess the magnitude of discretization effects in realistic 
computations.

In a finite volume the shifted partition function is given by 
\be
Z({\bz}) = \left(\prod_{\bp} \sum_{n_{\bp}} \right)
\exp\left[\sum_{\bp} n_{\bp}\left( - L_0\, \omega_{\bp} + i{\bp\cdot \bz} \right) \right]\, ,
\ee
where $\bp$ runs over the 3-dimensional momentum-space lattice 
$\bp_i =  2\pi \bm_i/L$ with i=1,2,3 and $\bm_i \in {\cal Z}$. 
As usual $n_{\bp}$ is the occupation number of the single-particle 
state labeled by $\bp$, see Ref.~\cite{Kapusta:2006pm}. By performing 
the geometric sums, the cumulant generator is given by
\be
\Fgen({\bz}) = \sum_{\bp} \ln{\frac{1-e^{- L_0 \omega_{\bp}+i{\bp\cdot \bz}}}
{1-e^{-L_0 \omega_{\bp}}}}\; . 
\ee
In the infinite volume limit 
\[
\sum_{\bp}\rightarrow V \int \frac{d^3 \bp}{(2\pi)^3}\; , 
\]
and rotational symmetry is restored. In the massless limit
$\omega_{\bp}=|\bp|$ and, by integrating over the norm of 
$\bp$ first, we obtain
\be\label{eq:genfct}
\frac{\Fgen({\bz})}{V} =
\frac{s}{4} \left[1- \frac{1}{(1+T^2 \bz^2)^2}\right]\, , 
\ee
where the entropy density is given by $s = 2\pi^2 T^{3}/45$. In the 
massive case $\omega_{\bp}=\sqrt{\bp^2+m^2}$, and we proceed by Taylor-expanding 
the logarithm and by performing the integration over the angular variables
\be
\frac{\Fgen({\bz})}{V} = \sum_{n\geq 1}
\frac{1}{2\pi^2 n}\int_0^\infty p^2 dp\, e^{-n L_0 \omega_{\bp}}
\left[1- \frac{\sin(n p|\bz|)}{n p |\bz|}\right]\, .
\ee
By also Taylor-expanding the sinus, and by using the integral representation of 
the modified Bessel functions $K_\nu(x)$ we obtain
\be\label{eq:genfct2}
\frac{\Fgen({\bz})}{V} = -
\frac{m^2 T}{\pi^{5/2}}   \sum_{j\geq 1}\frac{1}{j^2}
\sum_{n \geq 1} \frac{\Gamma(n+\frac{3}{2})}{(2n+1)!}\,
\left(- 2jm T \bz^2\right)^n\, K_{n+2}(j m/T) \,.
\ee
It is straightforward to verify that the cumulants 
generated from $\Fgen({\bz})$ in Eqs.~(\ref{eq:genfct}) and 
(\ref{eq:genfct2}) satisfy the recursive relation (\ref{eq:recK}).
For the massive case, the latter is implied by 
the recursive relation 
$\displaystyle \frac{d}{dx}\frac{K_\nu(x)}{x^\nu}=-\frac{K_{\nu+1}(x)}{x^{\nu}}$
among the modified Bessel functions.

\subsection{Free theory on the lattice}
Analogously to the continuum, the cumulant generator
of the scalar free theory discretized on the hypercubic 
in the standard way, i.e. with $\partial_\mu$ replaced by 
the forward finite-difference operator, is 
\be
\Fgen({\bz}) = \sum_{\bp}\ln
\frac{1-e^{-L_0 \omega_{\bp}+i{\bp \cdot \bz}}}{1-e^{-L_0\omega_{\bp}}}
= \frac{1}{2}\sum_{\bp}\ln
\frac{\cosh(L_0\, \omega_{\bp})-\cos({\bp \cdot \bz})}{\cosh(L_0\, \omega_{\bp})-1}\, ,
\ee
where 
\be
a\omega_{\bp} = 2\,{\rm asinh}\left(\frac{1}{2} a\sqrt{\hat \bp^2 +m^2}\right)\,,
\qquad
\hat \bp^2 = \frac{4}{a^2} \sum_{i=1}^3 \sin^2\left(\frac{a\, \bp_i}{2}\right)\; .
\ee
By expanding in the lattice spacing $a$, in the thermodynamic limit 
it is easy to show that discretization effects are well-behaved and 
scale as expected, i.e. proportionally  to $(a/L_0)^2$. In practice one can use 
$a\omega_{\bp}=2\ln u_{\bp}$,
$u_{\bp}= \frac{1}{2} a\sqrt{\hat \bp^2+m^2}+\sqrt{1 + a^2 (\hat \bp^2+m^2)/4} $,
so that $\cosh (L_0\, \omega_{\bp}) = \frac{1}{2}(u_{\bp}^{2L_0/a}+u_{\bp}^{-2L_0/a})$,
and carry out the sum over momenta numerically. 

\subsection{Scale-invariant theories\label{app:scc}}
The functional form in Eq.~(\ref{eq:genfct}) is more generally 
valid than just for the free massless theory. The combination of scale invariance 
and of the recursive relation (\ref{eq:recK}) fully fixes 
the cumulant generator $\Fgen({\bz})$. To show this we  
notice that scale invariance implies 
\be
k_{\{0,0,2n\}} = c_{\{0,0,2n\}}\, T^{2n+3}\,.
\ee
It is then not difficult to solve the recursion relation (\ref{eq:recK}) for $c_{\{0,0,2n\}}$
to find 
\be
k_{\{0,0,2n\}} = (n+1)\, (2n)!\, \frac{s}{4}\, T^{2n}\,.
\label{eq:kappa_conform}
\ee
For $\bz=\{0,0,z\}$ the cumulant generator can thus be written as 
\be
\Fgen(\{0,0,z\}) = V \sum_{n=1}^\infty \frac{(-1)^{n+1}}{(2n)!}\, k_{\{0,0,2n\}}\, z^{2n}\; , 
\ee
and by re-summing the series and using rotational invariance 
Eq.~(\ref{eq:genfct}) is obtained. A priori, one might have 
expected that in a scale-invariant 
theory, $\Fgen({\bz})$ can be an arbitrary function of $T|\bz|$. However, 
the Ward identities of energy and momentum conservation fix 
its functional form unambiguously.

\end{document}